\documentclass[journal]{IEEEtran}

\ifCLASSINFOpdf
\else
\usepackage[dvips]{graphicx}
\fi

\begin{document}

\title{Comparison of Ce$^{3+}$ and Pr$^{3+}$ activators in alkaline-earth fluoride crystals}

\author{E.Radzhabov and A.Nepomnyaschikh
\thanks{E.Radzhabov and A.Nepomnyaschikh are with Vinogradov Institute of Geochemistry, Russian Academy of Sciences, Favorskii street 1a,
P.O.Box 4019, 664033 Irkutsk, Russiaa.}
\thanks{Manuscript received April 19, 2005; revised January 11, 2007.}}


\maketitle

\begin{abstract}

The emission spectra of Ce$^{3+}$ or Pr$^{3+}$ doped CaF$_2$, SrF$_2$, BaF$_2$ excited by vacuum ultraviolet photons or by x-ray as well as excitation and absorption spectra in vacuum ultraviolet region (6-13 eV) were studied. The transfer of exciton energy is the main channel for Ce$^{3+}$ excitation in alkaline-earth fluorides. Three different stages of energy transfer were observed.  Pr$^{3+}$ excited by two processes, slow f-f luminescence excited by excitons, fast d-f luminescence excited by some fast process.

\end{abstract}


\section{Introduction}

\IEEEPARstart{I}{on} Ce$^{3+}$ still remains most popular activator for halide and oxide scintillators 
introducing both high efficiency of registration and fast decay time \cite{IEEEhowto:Krämer06}. Two
Ce$^{3+}$ emission bands are due to 5d-4f transitions having energy of transitions in fluorides near 4 eV.
Emission decay of Pr$^{3+}$ ions is  faster than that of Ce$^{3+}$ ions \cite{IEEEhowto:Rodny01}. Energies of several 5d-4f transitions of Pr$^{3+}$ are in 3.5-5.5 eV region. Other rare-earth ions show even faster decay, however energies of radiative transitions are shifted to vacuum ultraviolet region (above 6 eV), making difficult the detection of emission. Bromide and iodide matrices have the largest light yield, mostly due to smallest band gap. It seems that Ce$^{3+}$ ion only could effectively luminesce in these materials, while the emission of other rare-earth ions is quenched because the ground 4f$^n$ levels of these ions are in valence band \cite{IEEEhowto:Dorenbos07}. In this paper we compare the emission of Ce$^{3+}$ and Pr$^{3+}$ ions in alkaline-earth fluoride crystals. 

\section{Experimental}

Crystals CaF$_2$, SrF$_2$, BaF$_2$ doped by CeF$_3$ or PrF$_3$ were grown by Stockbarger method in graphite crucible in vacuum. Concentration of dopant varied from 0.005 to 10 molar percent. The excitation, absorption spectra in the energy range of 4-13 eV were measured with grating vacuum monochromator VMR2. The discharge hydrogen lamp VMF25 with MgF$_2$ window and dyoplasmatron were used as vacuum ultraviolet sources. Emission spectra were measured with grating vacuum monochromator VM2 (160-500 nm) and grating monochromator MDR2(200-900nm). Emission spectra were not corrected for spectral sensitivity response of monochromator and photomultipliers. Decay of emission was measured by digital oscilloscope 'Bordo-411' under excitation by 10ns pulse of x-ray tube 'Mira 2d'.Stationary emission was excited by Pd x-ray tube with 35 kV and 2-10 mA. 

At concentration less 0.3 mol.\% the samples were cutted from grown road. At higher concentration the sample were sawed from the road and one side polished. For simplicity we used a simple notation f-d or 4f-5d instead of full notation 4f$^n$-4f$^{n-1}$5d$^1$.

\section{Results}

\subsection*{Ce$^{3+}$ emission}

\begin{figure}[!t]
\centering
\includegraphics[width=3.5in]{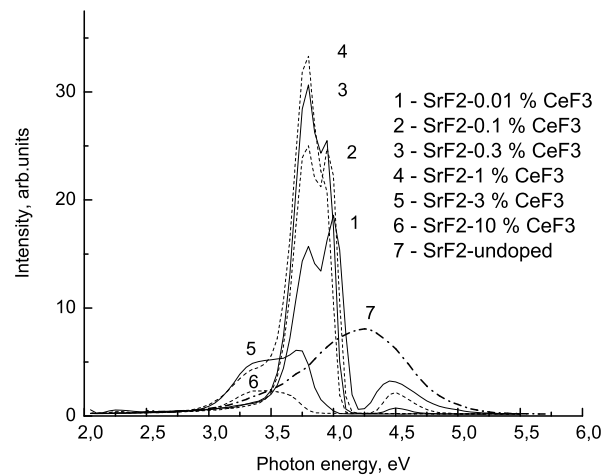}
\caption{Emission spectra of  SrF$_2$ crystals with different cerium
concentration at 295K under x-ray excitation. } 
\label{fig1}
\end{figure}

\begin{figure}[!t]
\centering
\includegraphics[width=3.5in]{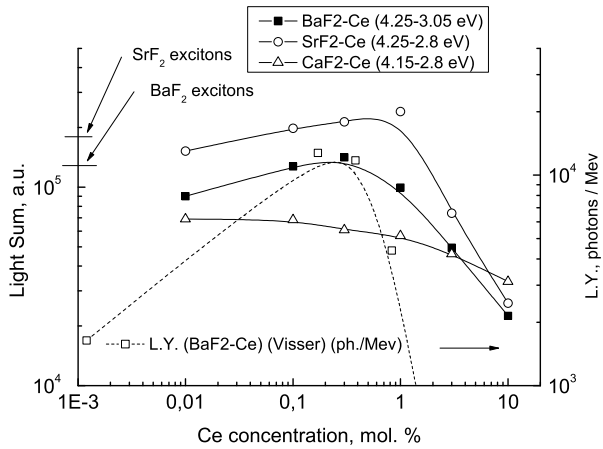}
\caption{Integral lightsum of Ce$^{3+}$ emission against of cerium concentration in
CaF$_2$, SrF$_2$, BaF$_2$ crystals.} 
\label{fig2}
\end{figure}

For all crystals the exciton luminescence continuously decreased with increasing of cerium concentration. Exciton emission becomes negligible near 1 mol.\% of CeF$_3$ (Fig.\ref{fig1}). Dependence of integral intensity of emission bands of Ce$^{3+}$ ions against of Ce concentration is shown on Figure. The highest light yield was found for SrF$_2$-1 mol. \% CeF$_3$ crystal.  The highest light yield of Ce$^{3+}$ emission for all crystals are close to initial light yield of exciton emission of undoped crystals (Fig.\ref{fig2}). The concentration of maximal light yield of BaF$_2$-Ce is the same as that observed in earlier paper \cite{IEEEhowto:Visser93}. It seems that the transfer of exciton energy is the main channel for Ce$^{3+}$ excitation. 

\begin{figure}[!t]
\centering
\includegraphics[width=3.5in]{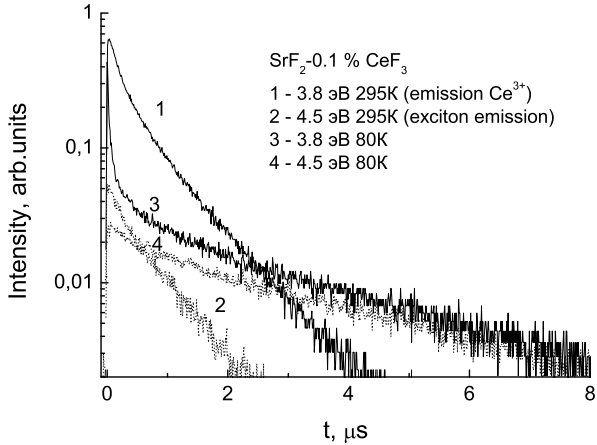}
\caption{Emission decay of SrF$_2$ 0.1 mol.\% PrF$_3$ crystals measured at 295K
and 80K. Sample was excited by 10 ns x-ray pulses.} 
\label{fig3}
\end{figure}

Emission decay of cerium ion doped into alkaline-earth fluorides under optical excitation into lowest energy absorption bands is equal about 30 ns \cite{IEEEhowto:Visser93}, \cite{IEEEhowto:Radzhabov04}, \cite{IEEEhowto:Wojtowicz00}. Under vacuum ultraviolet excitation into exciton and higher energies region the decay of BaF$_2$-Ce became unexponential \cite{IEEEhowto:Wojtowicz00}. Under x-pulses excitation the decay curves is also unexponential (Fig.\ref{fig3}). Decay curve can be fitted by few exponents, similar to exciton decay  \cite{IEEEhowto:Williams76}. With increasing of cerium concentration decay becomes faster. At low temperature the fast decay component becomes more pronounced (see Fig.\ref{fig3}). Whole decay curve can be described by at least three processes. Fast stage could be ascribed to resonance energy transition in nearest pairs of exciton and cerium ion. Middle stage up to microsecond range reduced by cooling of crystals (see Fig.\ref{fig3}) and could be ascribed to thermoactivated processes such as jumping of self-trapped excitons or holes. Slowest stage obviously connected with reabsorption of exciton emission over large distances.

\begin{figure}[!t]
\centering
\includegraphics[width=3.5in]{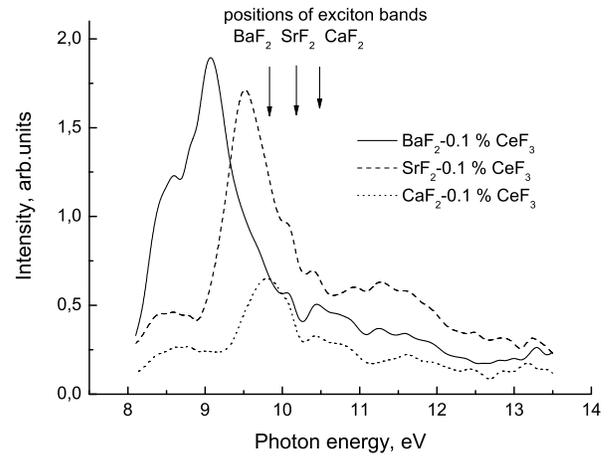}
\caption{Excitation spectra of Ce emission of CaF$_2$, SrF$_2$, BaF$_2$ crystals
doped by 0.1 mol.\% of CeF$_3$ at 295 K.  } 
\label{fig4}
\end{figure}

Excitation spectra of cerium emission show prominent peak at energies which is near 0.7 eV less than the exciton peak energies (Fig.\ref{fig4}). Similar peaks were observed for many activators in alkali halides. These bands usually assigned to excitons which were created in vicinity of activators.

\subsection*{Pr$^{3+}$ emission}

\begin{figure}[!t]
\centering
\includegraphics[width=3.5in]{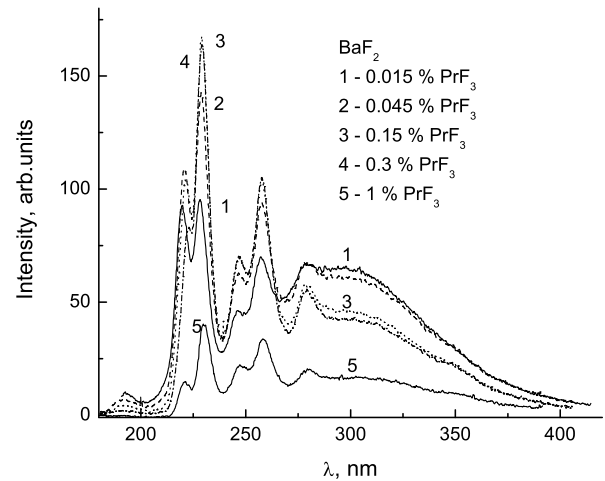}
\caption{Emission spectra of  BaF$_2$ crystals with different PrF$_3$ concentration
at 295
K under x-ray excitation. } 
\label{fig5}
\end{figure}

\begin{figure}[!t]
\centering
\includegraphics[width=3.5in]{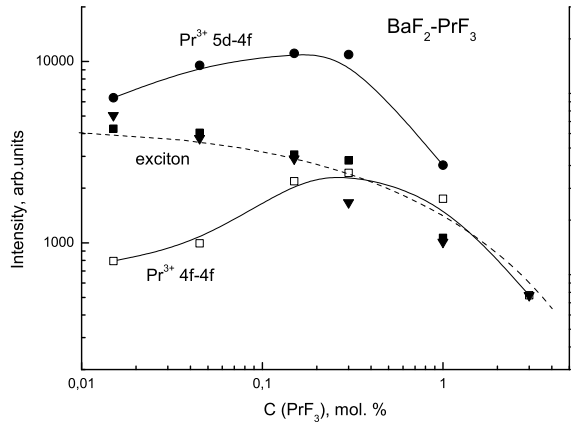}
\caption{Intensities of Pr$^{3+}$ 290nm(d-f), 500nm(f-f) bands and 290nm exciton
emission band against of PrF$_3$ concentration in BaF$_2$ crystals.} 
\label{fig6}
\end{figure}

Pr$^{3+}$ ions have two groups of emission bands. Bands 5d-4f are observed at 220-300 nm region (5.7-4 eV) (Fig.\ref{fig5}). Bands 4f-4f are observed at region larger 400 nm (less 3 eV). For characterization of relative intensity of both groups of lines we choose the maximal intensity bands at 230 nm (5d-4f) and 490 nm (4f-4f) (Fig.\ref{fig6}). With increasing of Pr concentration the intensity of exciton emission continuously decreased, however, in much slower rate in comparison with Ce doped samples. Maximal intensity of Pr emission was observed at 0.1-0.3 mol. \% of PrF$_3$ concentration (see Fig.\ref{fig6}). Concentration dependences of both groups lines are different (especially for CaF$_2$ and SrF$_2$), which points on different mechanisms of energy transfer. Relative intensity of 5d-4f emission against of exciton emission significantly decreased in a row of BaF$_2$-SrF$_2$-CaF$_2$.

\begin{figure}[!t]
\centering
\includegraphics[width=3.5in]{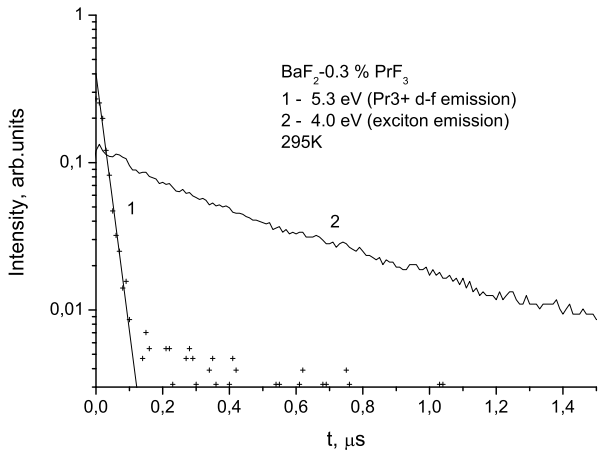}
\caption{Emission decay of BaF$_2$-0.3 mol.\% PrF$_3$ crystals measured  Pr d-f 
band and exciton band at 295
K under x-ray pulse excitation. } 
\label{fig7}
\end{figure}

Decay of 5d-4f Pr emission in BaF$_2$ is equal to near 27 ns (Fig.\ref{fig7}), decay do not depends on temperature and praseodymium concentration. The decay time is slightly longer than 22 ns measured under optical excitation in BaF$_2$-0.3 \% PrF$_3$ earlier \cite{IEEEhowto:Rodny05}. The fast decay of 5d Pr$^{3+}$ emission in several fluoride compounds was observed in 19-26 ns region \cite{IEEEhowto:Kück05,IEEEhowto:Makhov03}.  It seems the excitons have no influence on Pr 5d-4f emission.

\begin{figure}[!t]
\centering
\includegraphics[width=3.5in]{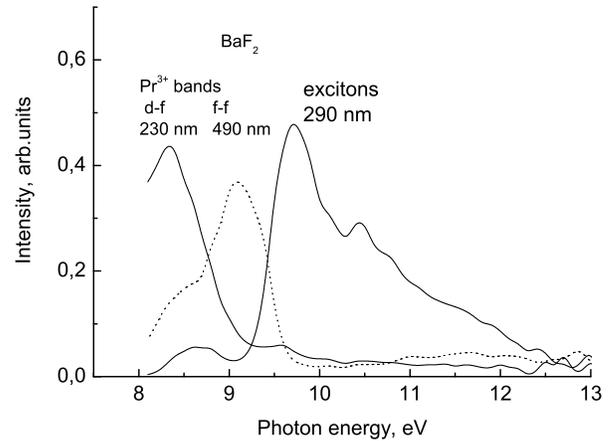}
\caption{Excitation spectra of BaF$_2$ -0.1 mol.\% PrF$_3$ crystals  at 295 K.
} 
\label{fig8}
\end{figure}

Excitation spectra for both Pr groups of transitions are different also (Fig.\ref{fig8}). Excitation spectra of f-f bands contains apparent peak on low energy side of exciton bands. Contrary to this the excitation spectra of d-f emission do not show any peaks above 4f-5d high energy band up to our limit at 13 eV (see Fig.\ref{fig8}). Evidently 5d-4f emissions are excited at higher energies.

\section{Discussion}

Excitation spectra of intraconfigurational f-f transitions of trivalent Nd, Er, Tm, Pr ions doped into several trifluorides were measured in early paper \cite{IEEEhowto:Yang78}. Excitation bands on low-energy wing of exciton bands were observed in all cases. Evidently excitonic mechanism of transfer energy to f-shell have common character. One can assume that the energy of self-trapped exciton (~4 eV in fluorides) transfer to appropriated excited f-levels of rare-earth ions. Evidently the exciton energy transfer have a low probability for Pr ions in alkaline-earth fluorides, it follows from weak exciton emission quenching by praseodymium ions (see Fig.\ref{fig5}). The exciton energy is not enough for excitation of 4f-5d transitions for all rare-earth ions but Ce$^{3+}$. 

The overlapping self-trapped exciton emission and Ce$^{3+}$ absorption, excitation spectra is high enough and above 1\% of CeF$_3$ all exciton energy is transfered to cerium ions (see Fig.\ref{fig2}). Because of slow decay of excitons and respectively cerium ions the alkaline-earth fluorides are not suitable for developing of fast scintillators.

The decay of Pr emission under x-ray pulses is fast as decay under optical excitation into f-d band. Intensity and decay time was not varied at temperatures from 80K to 300K (and even to 500K \cite{IEEEhowto:Shendrik10}). Fast d-f emission of BaF$_2$-0.3\%PrF$_3$ are excited above 18 eV, when excitation of core barium zone begins \cite{IEEEhowto:Rodny05}. In the case of BaF$_2$ and some other crystal the excitation of core band results in so-called crossluminescence, when electrons transfer from upper valence band to low-lying highest core band. Emission spectrum of crossluminescence in BaF$_2$ significantly overlap with absorption of Pr$^{3+}$ ions. Possibly it is the origin of highest BaF$_2$-Pr$^{3+}$ light yield among the other alkaline-earth fluorides, because in the CaF$_2$ or SrF$_2$ the crossluminescence is absent. The crossluminescent transition is absent in CaF$_2$, SrF$_2$ where the 5d-4f Pr luminescence observed also, but with lower intensity.

The results of excitation spectra measurements show clearly that in LiYF$_4$:Er$^{3+}$ the efficiency of the energy conversion from the matrix to Er$^{3+}$ d-f luminescence is extremely low in the region of fundamental absorption of the crystal till the energy of excitation reaches the threshold for processes of electronic excitations multiplication \cite{IEEEhowto:Makhov00}. This threshold lies in LiYF$_4$ : Er$^{3+}$ crystals at photon energy 22 eV, which is less than 2Eg (Eg=12.5 eV is the band gap for LiYF$_4$), and the effect is more pronounced at higher concentration of Er$^{3+}$. These features indicate that the mechanism of the energy transfer from the matrix to Er$^{3+}$ d-f luminescence in LiYF$_4$:Er$^{3+}$ is the so-called impact mechanism of electronic excitations multiplication when fast photoelectrons excite directly (by impact) emission centers (Er$^{3+}$ ions) in the crystal \cite{IEEEhowto:Makhov00}. This mechanism of excitation was thoroughly studied in thallium doped alkali chlorides \cite{IEEEhowto:Feldbach97}. This mechanism could operate in alkaline-earth flioride doped by praseodymium. In the case of BaF$_2$ the mechanism of excitation by hot photoelectrons can be reinforced by energy transfer from crossluminescent transitions. Future measurements of excitation spectra of CaF$_2$-Pr and SrF$_2$-Pr could refines the  details of energy transfer in Pr-doped alkaline-earth fluorides. 

BaF$_2$-PrF$_3$ is prospective media for developing of fast and thermal stable scintillator.

\section{Conclusion}
The transfer of exciton energy is the main channel for Ce$^{3+}$ excitation in
alkaline-earth fluorides. Three different stages of energy transfer were observed. Pr$^{3+}$ excited by two processes : slow f-f luminescence excited by excitons, d-f luminescence excited by some other fast process.

\section*{Acknowledgment}

This work was partially supported by grant 02-07-01057 from Russian Foundation for Basic Research (RFBR). The authors are grateful to V. Ivashechkin and V Kozlovskii for the growth of studied crystals.

\ifCLASSOPTIONcaptionsoff
  \newpage
\fi

\end{document}